\documentclass[preprint2]{aastex}

\slugcomment{Submitted to ApJ}

\shorttitle{The jet and the nucleus of 3C~270}
\shortauthors{Chiaberge et al.}

\begin{document}

\title{What do HST  and {\it Chandra} tell us about the  jet and the nuclear
region of the radio galaxy 3C~270?\thanks{Based  on observations obtained at
the  Space  Telescope Science  Institute,  which  is  operated by  the
Association of  Universities for Research  in Astronomy, Incorporated,
under NASA contract NAS 5-26555.}}


\author{Marco Chiaberge\altaffilmark{2}}
\affil{Space Telescope Science Institute, 3700 San Martin Drive,
Baltimore, MD 21218}
\affil{Istituto di Radioastronomia del  CNR - Via P. Gobetti
101, I-40129 Bologna, Italy}
\email{chiab@stsci.edu}

\author{Roberto Gilli}
\affil{Istituto Nazionale di Astrofisica (INAF) -- Osservatorio Astrofisico 
di Arcetri, Largo E. Fermi 5, I-50125 Firenze, Italy}
\author{F.~Duccio Macchetto\altaffilmark{3}, William~B. Sparks}
\affil{Space Telescope Science Institute, 3700 San Martin Drive,
Baltimore, MD 21218}
\author{Alessandro Capetti}
\affil{Istituto Nazionale di Astrofisica (INAF) -- Osservatorio  
Astronomico di  Torino,  Strada Osservatorio  20, I-10025   
Pino   Torinese,   Italy}


\altaffiltext{2}{ESA fellow}
\altaffiltext{3}{On assignment from ESA}


\begin{abstract}

The HST/STIS ultraviolet  image of the FR~I radio  galaxy 3C~270 shows
the presence  of a jet--like  structure emerging from the  position of
the  nucleus.   This feature,  which  represents  the first  jet--like
component ever  detected in the UV  in a radio galaxy  with jets lying
almost on the plane of the sky, has the same position angle as the jet
in the  radio and X--ray  images.  We propose two  different scenarios
for  the origin  of the  emission: i)  non-thermal synchrotron  from a
mildly relativistic component of the jet; ii) scattered light from the
nucleus,  where  a BL  Lac  source may  be  hosted.   Either of  these
pictures  would have  important consequences  for the  AGN unification
schemes  and for  our knowledge  of the  jet structure.   In  the {\it
Chandra}  image  a  faint   counter--jet  is  also  present.   From  a
comparative  analysis  of the  HST  images  and  {\it Chandra}  X--ray
spectrum, we find  that the nucleus is only  moderately obscured.  The
obscuring structure  might well reside in the  geometrically thin dark
disk  observed on large  scales.  This  fits the  scenario in  which a
standard  geometrically and optically  thick torus  is not  present in
FR~I radio galaxies.

\end{abstract}




\section{Introduction}

\begin{figure*}
\epsscale{2.3}\plottwo{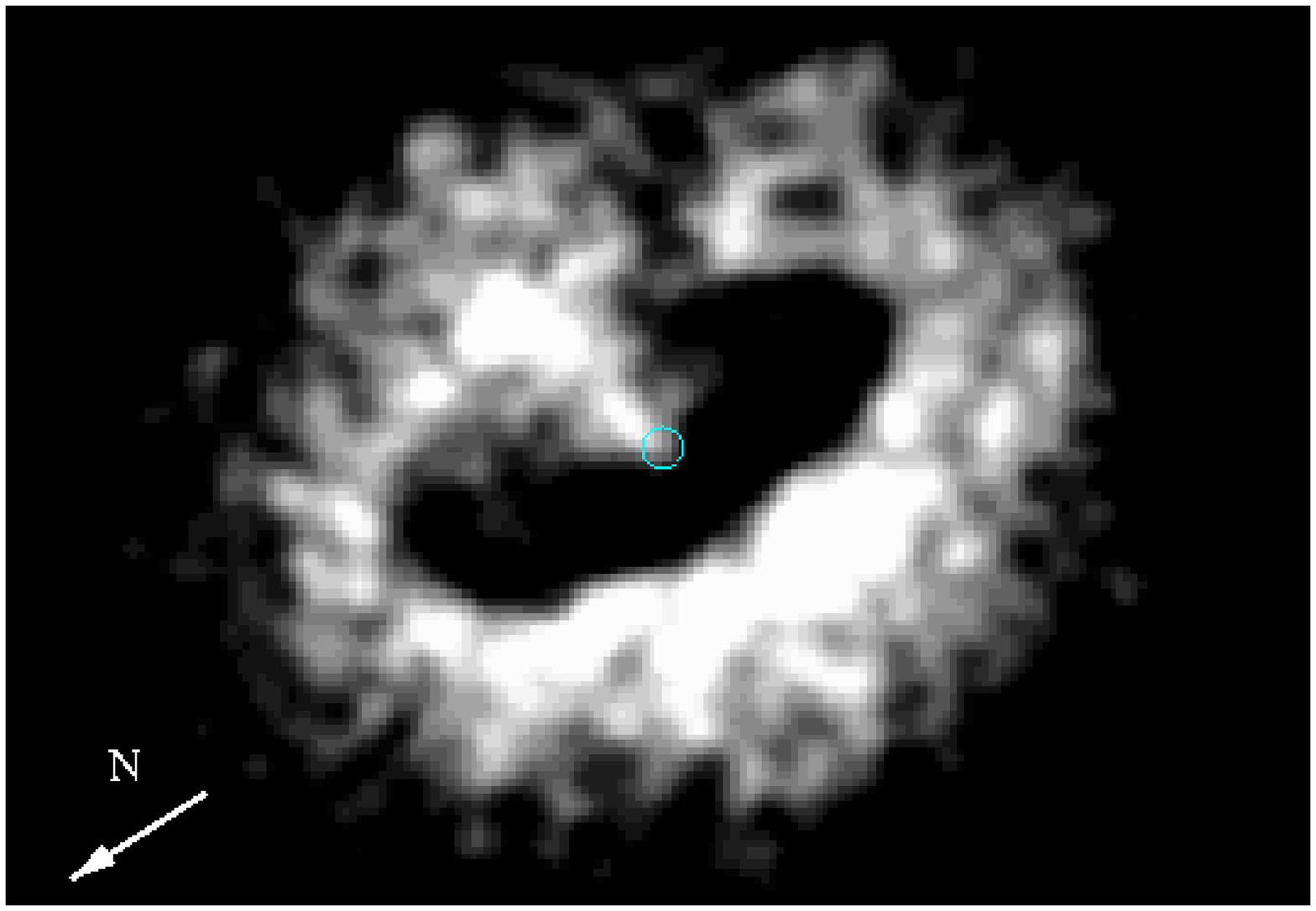}{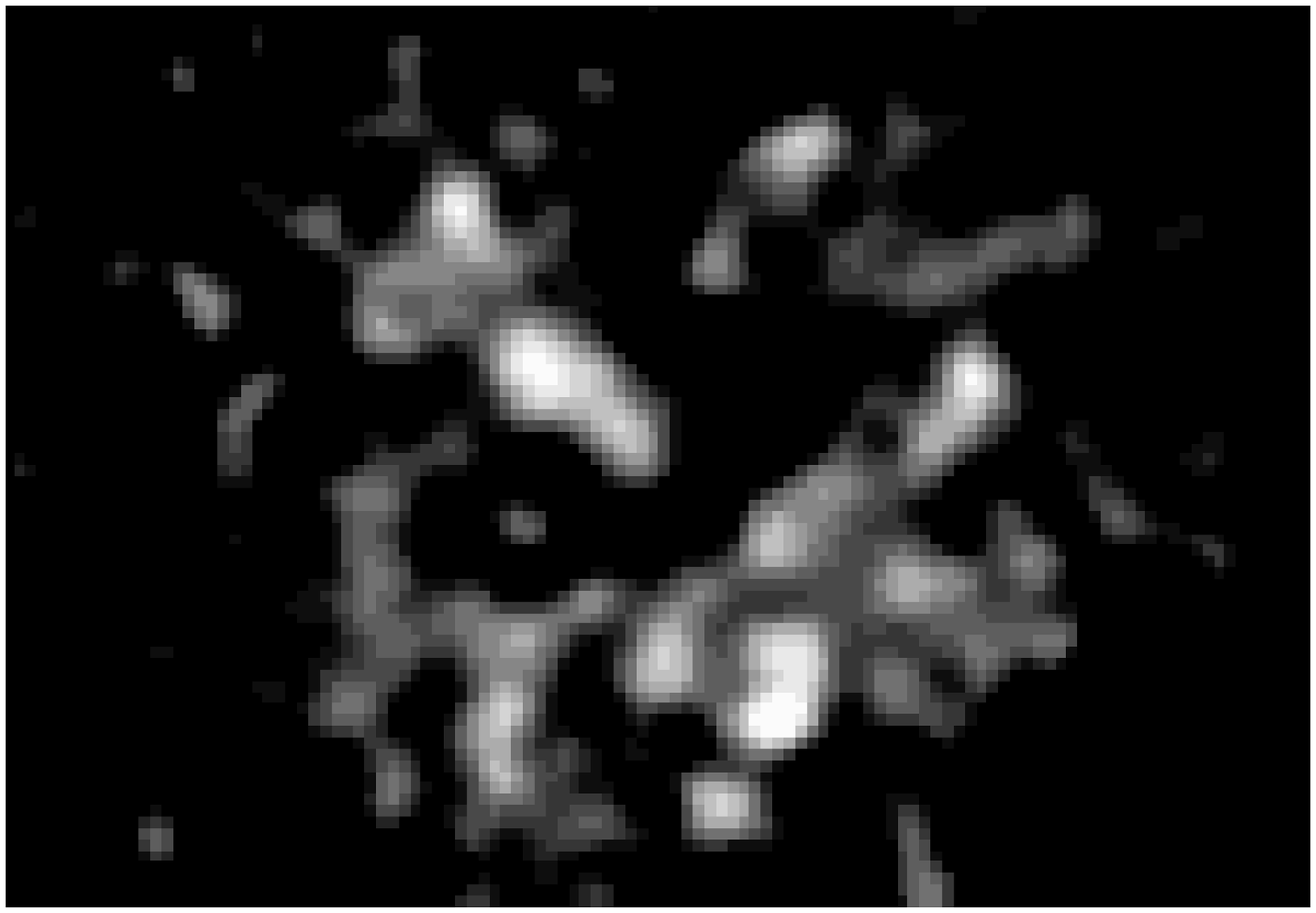}
\caption{Left     panel:    the     inner     regions    of     3C~270
($4.5^{\prime\prime}\times 3^{\prime\prime}$) as seen in the HST--STIS
UV  image, smoothed  with a  2 pixels  ($0.5^{\prime\prime}$) gaussian
kernel.  The  circle is centered at  the location of  the nucleus. The
``A''  component is  the  small jet--like  feature  emerging from  the
nucleus.  North  is indicated  by the arrow.   Right panel:  a ``color
image'' of the same region, obtained by dividing the UV by the V--band
image.  Both the  UV and the V--band images have  been smoothed with a
$0.1^{\prime\prime}$ gaussian kernel}
\label{im1}
\end{figure*}

The  unified scheme  for  low luminosity  radio--loud active  galactic
nuclei  associates low  luminosity FR~I  radio galaxies  with  BL Lacs
objects, or at least with most  of them.  According to the most recent
studies,  there is  growing evidence  that the  non--thermal radiation
produced at the base of the relativistic jet, on sub--pc scales, which
dominates  the overall emission  in BL  Lacs, is  also visible  in the
nuclei of their misaligned  counterparts. The nuclear sources observed
through  high  resolution HST  optical  and  UV  images in  the  great
majority  of  complete  samples of  FR~I  \citep{pap1,pap_uv,verdoes},
tightly correlate with the radio core emission, strongly arguing for a
common origin  (i.e non-thermal synchrotron emission from  the base of
the jet).   The discovery of  such non-thermal nuclei imply  that FR~I
radio  galaxies represent a  fundamentally different  manifestation of
the  AGN  phenomenon,  as  they  lack  at least  two  of  the  typical
characteristics of other  active galaxies: optically and geometrically
thick tori, and thermal disk emission.

On larger (kpc)  scales, jets have been observed  in the optical (e.g.
De Koff et  al.  1996, Martel et al.  1999, Capetti  et al.  2000) and
in the  UV band \citep{allen}, in  a substantial number  of FR~I radio
galaxies.  The origin for the emission of these jets is most plausibly
non-thermal   synchrotron   radiation.     Best   support   for   this
interpretation comes from detailed multiwavelength observations of the
different components of  the jets (e.g Perlman et  al. 2001, Sparks et
al.  1996), which show smooth spectra  from the radio to the UV.  
\cite{billjets} have shown  that optical jets in a  complete sample of
24 nearby 3CR FR~I are associated with sources in which dark disks are
seen  presumably  face-on.    Their  analysis  strongly  supports  the
scenario  in  which, as  a  result  of  relativistic beaming,  optical
non-thermal synchrotron emission from a jet is detected only if the jet
is   pointing  towards  the   observer,  forming   an  angle   to  the
line--of--sight $\theta < 30^{\circ}-40^{\circ}$.

On the other hand, observations  made with the {\it Chandra} satellite
have  shown that  X--ray jets  are common  among low  luminosity radio
galaxies  \citet{worrall2001}  and are  observed  in  quasars as  well
\citep{chartas,sambruna02}.   For  the   X-ray  emission  of  powerful
quasars'  jets,  the most  plausible  radiation  mechanism is  inverse
Compton  scattering  \citep[for  a  recent analysis  of  the  emission
processes  in X-ray  jets]{harris}. The  source for  the  seed photons
which are  upscattered to X--ray  energies might reside in  the Cosmic
Microwave  Radiation.   This  mechanism  requires high  velocity  bulk
motion  of the  jet  on the  kpc  scales, which  might  be present  in
powerful quasars  (and FR~II  galaxies) but is  not expected  in lower
luminosity  objects.  Another  possible source  for the  seed photons,
which  might well  apply to  low luminosity  radio galaxies'  jets, is
their  ``blazar'' nucleus.   \citet{cgc}  have shown  that, for  large
viewing angles to the jet, this  can be the dominant mechanism, if the
jet has a  slow velocity component (which might  be still relativistic
but  with a  Lorentz factor  of $\sim  2$). In  order  to discriminate
between the  above scenarios, a detailed  modeling of the  SED of such
large--scale  jets is  required.  Clearly,  this is  possible  only if
multiwavelength and  co--spatial data, from  radio to the  X-ray band,
are available.

The low luminosity  radio galaxy 3C~270 is an  ideal object to further
investigate  all  of  the  above  issues.  In  VLA  radio  maps,  this
nearby\footnote{We adopt a distance to 3C~270 of 41 Mpc \citep{faber}}
source  shows an  FR~I morphology  \citep{fr} with  symmetric  jet and
counter--jet  \citep{birkinshaw}.  VLBA  radio  images show  symmetric
jets on the  sub-pc scale, implying that the viewing  angle to the jet
is  quite large,  $>60^{\circ}$ \citep{jones97,jones00,jones01,piner}.
In the HST images,  the galaxy shows a large (hundreds--of--pc--scale)
obscure  dusty disk  \citep{ferrarese,martel}. The  jet  axis projects
onto the disk  minor axis and it is  plausibly almost perpendicular to
the  disk.  An unresolved  nucleus is  present in  the IR  and optical
\citep{fr1_sed,pap1},   but   it   is   not   detected   in   the   UV
\citep{allen,pap_uv}.

In this  paper we show  that the HST/STIS  UV image of 3C~270  shows a
further feature, which  is possibly connected to the  jet.  As we will
show  in the following,  such feature  is unexpected  in an  object as
3C~270, in  which the jets  almost lie on  the plane of the  sky.  The
presence  of this  component, together  with both  the absence  of the
nucleus in  the UV and the  information from the  {\it Chandra} X--ray
data, give us a unique global picture of the nuclear structure of this
object. The  X--ray image also shows  the presence of a  jet and, more
surprisingly,  of a  counterjet. All  of these  issues  have important
implications for the AGN unified models.

This paper  is organized as follows: in  Section \ref{observations} we
describe  the HST  and  {\it  Chandra} observations  and  we show  the
results;  in Section \ref{discussion}  we discuss  our results  and we
propose different scenarios which can account for the observations; in
Section \ref{conclusions} we present a  summary of our findings and we
draw conclusions.

\section{Observations and results}
\label{observations}

\subsection{HST observations}

The UV HST-STIS  image was taken as part of the  UV Snapshot survey of
3CR  radio  galaxies  \citep[SNAP  8275]{allen},  using  the  NUV-MAMA
detector  and  the  F25SRF2  broad  band  filter,  whose  transmission
function  is centered  at  2320  \AA.  The  image  has been  processed
through  the HST  pipeline calibration.   IRAF standard  packages have
been used  for the analysis.   In Fig.~\ref{im1} (left panel)  we show
the  central $4.5^{\prime\prime}\times  3^{\prime\prime}$ of  the STIS
image, which has been smoothed by convolving the original image with a
2 pixels  (corresponding to $0.048^{\prime\prime}$)  sigma gaussian in
order to improve the rather low signal--to--noise ratio ($S/N \sim 10$
in the  center of the  galaxy).  The most  apparent feature in  the UV
image is  the $\sim$ three hundred pc--scale  dark disk \citep{jaffe}.
Emission  from  the host  galaxy  stellar  component  is also  clearly
visible.   The circle  represents the  position of  the center  of the
galaxy, which has been identified by fitting ellipses to the isophotes
of the  galaxy, in the region  outside the dark disk,  between 1.5 and
2.5   arcsec   from  the   center.    Due   to   the  relatively   low
signal--to--noise ratio of the UV image, the position of the center of
the galaxy is known with an uncertainty of $\sim \pm 1$ pixel.

While  an unresolved  nucleus  is present  in  the near--infrared  and
optical HST images \citep{ferrarese,fr1_sed}, in the UV image a faint,
elongated, jet--like  structure appears to  emerge from the  center of
the galaxy, and the nucleus is not detected \citep{allen}.  As pointed
out by \citet{ferrarese}, the nucleus is probably not at the center of
the  galaxy isophotes.   However, the  error  on the  position of  the
center in the UV  image is of the same order as  the shift between the
isophote  center and  the position  of the  unresolved nucleus  in the
optical images.  Therefore such small shift ($0.023 \pm 0.009$ arcsec)
is not significant for  our purposes.  

The apparent angular dimensions of  the jet--like feature (to which we
will  refer in the  following as  the ``A''  component) are  $\sim 0.3
\times  0.1$ arcsec,  which  correspond  to $\sim  60  \times 20$  pc.
However, the width of the feature is consistent with being unresolved.
The  ``A''   component  lies  at   the  same  position   angle  ($\sim
90^{\circ}$) as the  jets observed in the radio, both  on the VLBI and
VLA scales (Jones et al.  2000, Birkinshaw \& Davies 1985), and in the
X--ray band (see below).  Clumps of emission, possibly superimposed on
other components aligned  with the jet, are present  on larger scales,
across the western edge of the obscure disk, up to a distance of $\sim
0.8$ arcsec from the location of the nucleus.

We have  measured the flux of  the ``A'' component,  by extracting the
count rate  from a  box of  $13 \times 7$  pixels, and  evaluating the
background by  averaging the counts  in the pixels contiguous  to such
box.  The background evaluation is clearly the largest source of error
in the process of evaluating  the flux of this component. Adopting the
PHOTFLAM  parameter  in  the  image header\footnote{The  HST  internal
calibration is  expected to  be accurate to  $5\%$.}  to  convert from
counts  to flux density,  we obtained  $F_\lambda =  5 (\pm  2) \times
10^{-19}$  erg  cm$^{-2}$ s$^{-1}$  \AA$^{-1}$.   At  the distance  to
3C~270 this corresponds to a  luminosity $\lambda L_\lambda = 2.3 (\pm
1.0) \times 10^{38}$ erg s$^{-1}$.

\subsection{Chandra observations}

\begin{figure}
\epsscale{1} \plotone{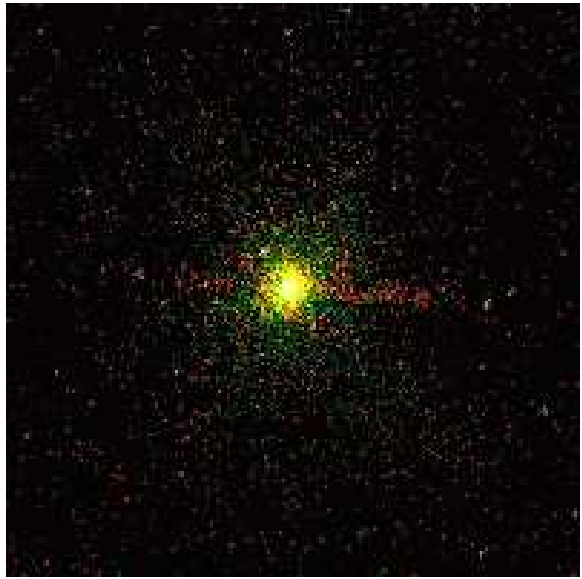}
\caption{The {\it Chandra} X--ray image  of 3C~270. The image scale is
100 arcsec $\times$ 100 arcsec. North is up, East is left.}
\label{chandra}
\end{figure}

\begin{figure}
\epsscale{1} \plotone{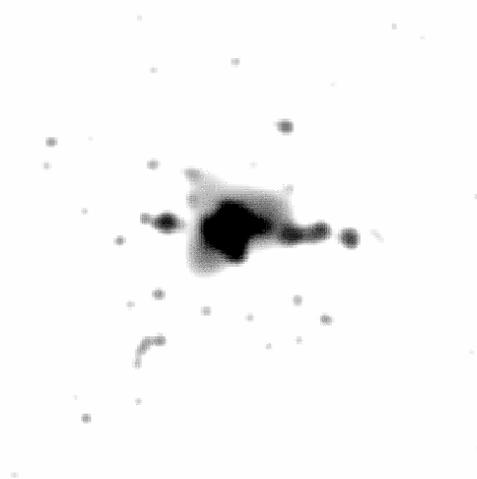}
\caption{Soft  X--ray  (0.3--0.6 keV)  adatptively  smoothed image  of
3C~270. The image scale and orientation are as in Fig. \ref{chandra}.}
\label{softx}
\end{figure}

We analyze  a {\it Chandra} archival  34 ksec observation of  3C~270 (P.I.
Birkinshaw), obtained on  2000 May 6 with the  back illuminated ACIS-S3
chip. Since significant  pileup effects were expected,
a  subarray  configuration  was  used,  which allowed  to  reduce  the
frametime to 1.8 sec and the estimated pileup fraction to $\sim 10\%$.
The data reduction was performed  with CIAO v2.2 applying the standard
corrections  for bad and  flaring pixels,  and filtering  for standard
ASCA grades. An updated gain map included in the CALDB 2.8 release was
used, which reduces the  calibration uncertainties at low energies. We
also eliminated about  12\% of the exposure time  where the background
flux was $3\sigma$ above the average value.

The  X-ray  image (Fig.   \ref{chandra})  clearly  shows the  nucleus,
diffuse  emission  from the  hot  corona,  a  kpc-scale jet  (pointing
westwards) and,  surprisingly, a  faint counterjet. The  counterjet is
clearly visible in the X--ray  image when only the soft (0.3--0.6 keV)
is displayed  (Fig.  \ref{softx}).  The jet--counterjet  flux ratio is
$2.2\pm 1.1$,  in substantial agreement  with what is observed  in the
radio band \citep{xu,piner}.

To avoid  contamination from  the kpc-scale jets  and thermal  gas, we
extract the  nuclear spectrum  in a circle  with aperture radius  of 4
pixel ($\sim  2$ arcsec), which is twice  as large as the  FWHM of the
observed Point  Spread Function.  The background  spectrum is measured
in an external  region on the same chip.  The  background flux is less
than 1\% of the source flux,  therefore our results are not altered by
measuring the background spectrum in a different region.

We  carry   out  the  spectral   analysis  using  the   XSPEC  v11.1.0
package. 
We fitted the 0.3--9 keV  spectrum of 3C~270 with an absorbed powerlaw
(the AGN)  plus a  thermal plasma ({\tt  mekal} model in  XSPEC).  
Since it is  likely that the observed spectrum  has been significantly
hardened by pileup, we convolve  the model with the {\tt pileup} model
in XSPEC \citep{davis01}.
In   addition   to   the   thermal   component  we   obtain   a   hard
power--law\footnote{The spectral  index $\alpha$ is  defined by $F_\nu
\propto  \nu^{-\alpha}$} ($\alpha=0.30\pm0.06$)  which  is similar  to
what     has    been     found    in     ASCA    and     ROSAT    data
\citep{matsumoto,sambruna99,fr1_sed}.

\begin{figure}
\epsscale{1}\plotone{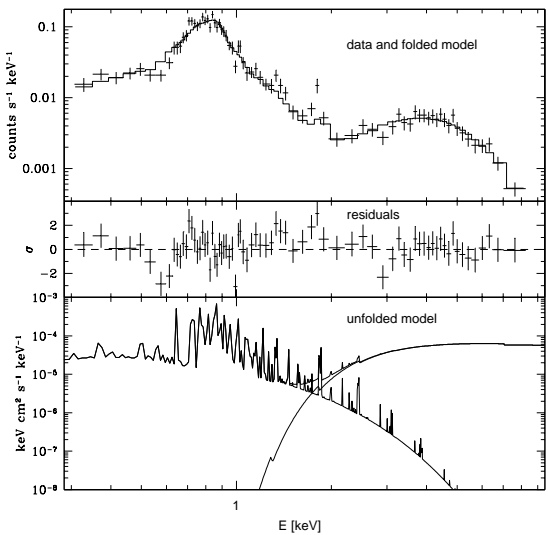}
\caption{The X-ray spectrum and model for 3C~270.}
\label{specx}
\end{figure}

Most   importantly,  and   differently   from  the   ROSAT  and   ASCA
observations,  we  find evidence  for  a  moderate  amount of  nuclear
absorption ($N_H =  6.33\pm0.78\times 10^{22}$ cm$^{-2}$).  The X--ray
spectrum of 3C~270  is shown in Fig. \ref{specx}.   

Note  that the presence  of moderate X--ray absorption  is rather
common among FR~I. \citet{sambruna99}  studied nine FR~I galaxies with
the ASCA satellite,  finding a broad range of  absorbing columns.  The
typical  values   for  $N_H$  lie  between   $10^{21}$  and  $10^{22}$
cm$^{-2}$,  and for  three  out of  the  nine FR~I  galaxies in  their
sample, they find no intrinsic absorption.  Very recent work with {\it
Chandra} (e.g.  Hardcastle et al.  2001, 2002) basically confirms this
picture.  In order  to check for consistency with  the previous X--ray
observations of  3C~270, we  have reanalyzed the  ASCA data.   We find
that a  satisfactory fit can  be obtained by setting  the parameters
for the power-law  component and absorption to those  we have obtained
from the Chandra observation.

We also  find that  a $\sim  6.4$ keV iron  line is  not statistically
required by our fit. However, its  presence cannot be ruled out by the
{\it Chandra}  data, since  we have no  stringent upper limits  on its
equivalent width (380  eV at 90\% c.l.). In  Table \ref{chandrafit} we
summarize the  parameters for  the best fit  ($\chi^2/dof=95/69$), and
fluxes and  luminosities of the different  spectral components (errors
are at 90\% c.l.).  Note that  the 2--10 keV flux is comparable to the
ASCA value,  while the  {\it Chandra}  0.5--2 keV flux  is a  factor 4
lower, since,  due to the the  small extraction area we  adopt, we are
missing most of the diffuse soft thermal emission.  Instead, the total
2--10 keV flux is dominated by the (nuclear) power--law component.  By
subtracting the thermal component from  the total flux, we obtain that
the observed flux  of the power--law component in  the range 2--10 keV
is $6.88 \times 10^{-13}$ erg s$^{-1}$ cm$^{-2}$.

\begin{deluxetable}{l c c c c}
\tablewidth{0pt}   \tablecaption{Best  fit   parameters,   fluxes  and
luminosities in the  two energy ranges $0.5-2$ keV  and $2-10$ keV for
the X-ray spectrum of 3C~270.  The Galactic column density is fixed to
$1.87\:10^{20}$ cm$^{-2}$ \citep{stark}.}   \tablehead{ \colhead{~ } &
\colhead{Total}   &    \colhead{thermal}   &   \colhead{Power-Law}   &
\colhead{P-L (de--absorbed)\tablenotemark{c}} }

\startdata
$F_{0.5-2}$\tablenotemark{b}  &  1.29    & 1.26 & --            &       --    \\
$L_{0.5-2}$\tablenotemark{c}  &  0.26    & 0.25 &  --           &       --    \\
$F_{2-10}$\tablenotemark{b}   &  6.91    & --   &   6.88        &      9.58   \\
$L_{2-10}$\tablenotemark{c}   &  1.38    &      &   1.37        &      1.91   \\

\tableline
\tableline
\\
P--L spectral index $\alpha$   &  $N_H$\tablenotemark{d}   &  kT               &   $Z/Z_{\odot}$  \\
\tableline
$0.30 \pm 0.06$   & $6.33\pm 0.78$ & $0.61(\pm 0.04)$ keV & $0.48 \pm 0.07$ \\
\enddata
\tablenotetext{a}{The
de--absorbed flux and luminosity are calculated assuming the value for
$N_H$ as obtained from the fit.}
\tablenotetext{b}{In units of $10^{-13}$ erg cm$^{-2}$ s$^{-1}$}
\tablenotetext{c}{In units of $10^{41}$ erg s$^{-1}$} 
\tablenotetext{d}{In units of $10^{22}$ cm$^{-2}$}
\label{chandrafit}

\end{deluxetable}

\section{Discussion}
\label{discussion}

Two are the most important  questions arising from the analysis of the
STIS image of the nuclear regions of 3C~270:

$\bullet$ what is the nature of the ``A'' jet--like feature?

$\bullet$ what is the role of nuclear absorption in this radio galaxy?

In  the following  sections, we  test whether  the available  data are
consistent with the synchrotron scenario for the jet-like feature, and
we  discuss the  implications  for  the AGN  unified  model. Then,  we
consider  a  different  scenario  for the  observed  emission  (namely
scattered  radiation  from the  ``blazar''  nucleus)  which has  again
important  implications  for  the  unification schemes.   Finally,  we
discuss  the  role  of  nuclear  absorption,  and  we  suggest  future
observations that  may offer  a definitive discrimination  between the
two.

\subsection{A UV jet?}

A fundamental tool  in order to constrain the  emission process of the
jet--like structure  in 3C~270 is  the information about  its spectral
shape.   Unfortunately, the  ``A'' component  is not  detected  in any
other HST image, and  no co--spatial observations are available either
in the radio or in the  X--ray bands.  

The absence  of a corresponding feature  in the optical  images is not
surprising: we can predict the  expected counts of such feature in the
optical (F547M)  image, by considering  the flux density  of component
``A''  in the  UV.  Assuming  a  spectral index  $\alpha=1$, which  is
common among  synchrotron emitting  sources in the  optical-UV regime,
the expected flux  is $F_{5500} \sim 3 \times  10^{-19}$ erg cm$^{-2}$
s$^{-1}$ \AA$^{-1}$.   Using the WFPC2 exposure  time calculator, this
translates in a count  rate of $4\times 10^{-2}$ counts/sec integrated
on an area of $13 \times  7$ pixels, corresponding to an average value
of    0.4   counts/pixel    in    the   F547M    image   (which    has
$t_{exp}=800$s). This  value is less  than 0.3\% of the  background in
the  region of  the  dark disk,  at  the position  angle  of the  jet.
Therefore if  the ``A'' component is  a synchrotron jet,  it cannot be
detected in the available optical images.

In Fig.  \ref{im1} (right panel)  we show a ``color'' image UV--V.  We
have chosen  the V--band image since  in the other  optical images the
PSF  from the nuclear  component dominates  the central  regions.  The
galaxy emission fills the field of view of both instruments, therefore
the background is undetermined and  the estimate of the color index of
the jet is not  straightforward \citep{billcolor}. Such an analysis is
well  out of  the  purposes  of this  paper.   However, a  qualitative
analysis  of this  image  tells us  that  i) the  region with  similar
spectral  properties  (the  jet?)   extends  further  than  the  ``A''
component,  and its  color  is almost  constant  on a  scale of  $\sim
0.8^{\prime\prime}$  ii)  the ``jet''  is  bluer  than the  underlying
stellar emission of the host galaxy.

It is  also important  to compare  the UV to  the radio  emission.  As
already mentioned, unfortunately no co--spatial radio observations are
available in the literature.  However, since the distance in frequency
between radio  and UV measurement  is relatively high, we  can roughly
estimate the radio  flux by subtracting the VLBI  total flux (observed
on the  scale of  $\sim 50$ mas  from the  nucleus) from the  VLA core
flux.  This  results in a rough  estimate of the  emission coming from
the region  between the VLA and  the VLBI scales, i.e.  some tenths of
arcsec.  Adopting the fluxes given by \citet{xu}, we obtain a value of
the radio--UV spectral index $\alpha_{\rm r-UV}\sim 0.8-0.9$, which is
compatible with other synchrotron emitting jets \citep{billm87}.  Note
that due  to the large difference  in frequency between  the radio and
the UV, an error of a factor  two on one of the two measurements would
result in a variation of only $\Delta \alpha_{\rm r-UV} = 0.05$.

\subsection{Synchrotron emission}

The detection  of synchrotron emission from  a jet oriented  at such a
large angle to  the line--of--sight is not expected  at either optical
or UV wavelengths,  since the observed flux from  relativistic jets is
certainly affected by relativistic beaming.  The observed flux is $F =
\delta^{4}  F^{\prime}$, where  $\delta=1/\Gamma(1-\beta \cos\theta)$,
$\Gamma$ is  the bulk Lorentz factor, $\beta=v/c$  and $F^{\prime}$ is
the flux  observed in  the comoving  frame of the  jet.  3C~270  has a
quasisymmetric jet  and counter--jet  (on all scales  and both  in the
radio and in  the X--ray band), therefore, in  the standard picture,
the  viewing angle should  be quite  large.  This  results in  a small
value for  $\delta$, and  most of the  jet radiation should  be beamed
away from our line--of sight\footnote{For a jet bulk Lorentz factor of
10, $\delta(60^{\circ})= 0.2$, which would reduce the observed flux by
a factor $\sim 600$, while for $\Gamma=5$ the flux is still reduced by
a  factor $\sim  40$.}.  Although  the  extreme Lorentz  factors
($\Gamma  ~10-20$)  required  in  order  to account  for  the  BL  Lac
phenomenon are  probably limited to  the very innermost region  of the
jet (sub--pc),  it is not yet  clear where the  jet deceleration takes
place and what are the characteristics of the velocity gradients along
the jet.  \citet{giovannini} and  \citet{xu} have pointed out that the
low  number of  objects  with detected  VLBI  counterjets in  randomly
oriented samples implies  that motions with $\Gamma \sim  3-5$ must be
present, at least  on the pc scale. 

On  larger   scales,  the  study  of  jet   asymmetries  performed  by
\citet{laing99} suggest  that FR~I jets  are still relativistic  up to
$\sim   1$  kpc  from   the  core.    Furthermore,  the   analysis  of
\citet{billjets}  of a  statistically complete  and  randomly oriented
sample  of nearby  FR~I  from the  3CR  catalog, for  which  a set  of
homogeneous  HST observations  is available,  shows  that relativistic
beaming  plays indeed  an  important  role in  the  detection of  FR~I
optical jets on the kpc scale.


The chance  of detecting a UV  synchrotron jet in  a highly misaligned
source  such  as 3C~270  is  significantly  increased  if lower  speed
components are present  in the jet.  This scenario  might find support
from radio observations of this object.  \citet{piner} have calculated
that the viewing  angle of the 3C~270 jet is  $\sim 63^{\circ} \pm 3$,
on the basis of the jet-counterjet ratio and the apparent jet velocity
from  VLBI  data.   According  to their  observations,  the  intrinsic
velocity  is $\beta=0.46  \pm0.02$.  Although  their estimate  is only
based  on the  motion of  a single  local maximum  on the  jet profile
plots, this would  corresponds to a mildly relativistic  motion with a
Lorentz factor $\Gamma=1.13$, which might be expected if the jet has a
velocity structure on  the pc (and sub-pc) scales,  for example a fast
spine and a slower  layer \citep{laing99,pap3}\footnote{If we assume a
value of $\Gamma  \sim 3-5$ for the jet velocity,  which appears to be
more typical of the broad  class of low luminosity radio galaxies, the
jet-to-counterjet  ratio (in  the radio  band and  on the  VLBI scale)
implies that the viewing angle can be as large as $\sim 80^{\circ}$.}.
In  this  case,  the  detection  of  synchrotron  emission  in  highly
misaligned sources ($\sim 60^{\circ}$ or more) is possible, since with
such  low  Lorentz  factors  the  de-beaming effect  does  not  affect
dramatically the  jet radiation.  Note  however, that if all  FR~I had
significant emission  from such a  slow component, we would  expect to
observe optical  jets in  most of  them, which is  not the  case.  The
probability of  observing them  might increase in  the UV, due  to the
enhanced contrast between the  jet emission and the underlying stellar
galaxy component.  However, HST/STIS UV images are available for 14 of
the 24  FR~I of the  \citet{billjets} sample.  With the  only possible
exception   of  3C~270,   no   ``new''  jets   have  been   discovered
\citep{allen}.  This further supports  the idea that detection of jets
is  more  strongly determined  by  relativistic  beaming  than by  the
contrast to the host galaxy emission.

We stress that whatever the viewing angle to 3C~270 (which however has
to be rather large), the detection of a synchrotron UV jet implies that
{\it we are  observing a component of the jet  which moves slower than
what is expected for a typical  BL Lac nucleus}, for which $\Gamma$ is
constrained  to be $\sim  10-20$, at  least at  optical-UV frequencies
(e.g. Tavecchio  et al.  1998,  Sikora et al. 1994).   A low--velocity
component  must  be present  on  larger  scales,  to account  for  the
observation  of   X--ray  emission   from  the  kpc--scale   jet  (and
counter--jet).

We can further  test this scenario through a  reverse argument, in the
frame of the AGN unification schemes.  We can calculate the luminosity
of the  ``A'' component  as seen  from an observer  placed at  a small
angle  to the  axis of  the jet  ($\theta \sim  1/\Gamma$),  under the
assumption of highly  relativistic motion of such a  component.  If we
assume $\Gamma  =15$ for  the bulk  Lorentz factor of  the jet,  for a
large viewing angle ($60^{\circ}-80^{\circ}$) we obtain $\delta_A \sim
0.1$.   Therefore, the  luminosity  $\nu L_\nu$,  as observed  on-axis
would be $L_{on-ax}\sim  (L_A/\delta^4_A) \delta_{15}^4 = 10^{47}$ erg
s$^{-1}$.  We can compare this to the optical luminosity ($\nu L_\nu$)
of a  BL Lac  which has  the same extended  (unbeamed) radio  power as
3C~270 ($\log L_{1.4} =31.4$  erg s$^{-1}$ Hz$^{-1}$).  Although there
is a large scatter in the data ($\sim \pm 1$ dex), an average value of
$L_{BL}=5\times  10^{44}$   erg  s$^{-1}$  (see   e.g.   Chiaberge  et
al. 2000)  represents a good estimate for  its luminosity.  Therefore,
$L_{on-ax}$ is  at least 2  orders of magnitude higher  than expected.
In order for the ``A'' component  not to outshine the inner BL Lac, we
must set an  upper limit for its bulk Lorentz factor  $\Gamma < 7$ (if
the viewing  angle to the jet  in 3C~270 is  $\theta ~60^{\circ}$, and
even lower if $\theta$ is  larger). This limit is therefore compatible
with the  scenario in which the  UV emission observed on  the scale of
tens--of--pc is produced by a rather slow component of the jet.

\subsection{Scattered emission from a misoriented BL Lac}

Considering the  potential difficulties about  the synchrotron origin,
we explore  an alternative  scenario, namely scattered  radiation from
the central regions.  Such a  phenomenon might indeed be observed {\it
only}  in highly  misoriented sources.   The observation  of scattered
radiation  (both in  the form  of continuum  and emission  lines) from
absorbed nuclei  in Type 2 AGNs  is the most  striking confirmation of
the unification schemes.   The best examples of such  a process can be
found  in spectropolarimetric  data of  Seyfert 2  and  radio galaxies
\citep{ski85,  cohen} and  in  HST polarimetric  images  of Seyfert  2
galaxies \citep{capetti95,kishimoto}.   The observed phenomenology has
been  successfully interpreted  as due  to  the presence  of a  hidden
quasar  nucleus,  in  which  both  the  continuum  emission  from  the
accretion disk  and the  broad emission line  region are  seen through
scattered  light.  The  nature of  the ``mirrors''  is  still unknown,
since  the  ambiguity   between  electron  scattering  and  dust-grain
scattering has not been clearly solved.

Since  there are  strong  clues  that the  nuclear  emission from  any
accretion disk is rather  low in FR~I \citep{pap1,skim87,perlman}, the
dominant nuclear  source for the  scattered light might reside  in the
base of the relativistic jet.  The material responsible for scattering
might be  either the ISM or the  jet itself.  In the  former case, the
observed  structure (i.e.  component ``A'')  represents a  {\it direct
evidence of  the jet beam}.   Note that in  such a scenario,  the fact
that the width of component A is unresolved at a distance of $\sim 60$
pc from the nucleus sets a limit to the half opening angle of the beam
of the inner misoriented BL Lac as being $\theta < 9.5^{\circ}$.  This
in turn translates  to a bulk Lorentz factor of the  jet $\Gamma > 6$,
which is  consistent with  the highly relativistic  velocities $\Gamma
\sim  10-20$ required  by  modeling  of the  BL  Lac overall  emission
\citep{gg98}.

Let us now calculate the  particle density of the scattering material.
The  radiation  observed  by  any  material located  on  the  line  of
propagation of the jet can be estimated by considering the emission of
a  BL Lac  object of  the same  intrinsic total  power as  3C~270.  We
assume  again  $L_{BL}=5\times   10^{44}$  erg  s$^{-1}$.   Since  the
luminosity of component ``A'' is  $L_A = 2.3 (\pm 1.0) \times 10^{38}$
erg s$^{-1}$  and it has  a dimension of  $\sim 20 \times 60$  pc, the
particle   density  inside   the   scattering  cloud   using  $L_A   =
(L_{BL}/(4\pi R^2)) \times  \sigma_p \times V \times n_p$,  where R is
the distance to the source  of radiation ($R\sim 50$pc), $\sigma_p$ is
the particles cross section, $n_p$  is the particle density and $V$ is
the  volume  of  the  scattering  region. In  case  of  scattering  by
electrons  we  obtain $n_e  \sim  0.4$  cm$^{-3}$,  while in  case  of
scattering by dust we have  (assuming normal dust/gas ratio) $n_H \sim
3 \times 10^{-4}$ cm$^{-3}$.  More precisely, these values should
be considered as lower limits,  since the width of the ``A'' component
might be unresolved, and therefore the volume of the jet might in turn
be  overestimated.  Although  both of  them appear  to  be reasonable
values, we stress that from the normalization of the thermal component
in the X--ray spectrum\footnote{  The normalization of the thermal
component is defined as $N = (10^{-14} \int_V n_e n_H dV)/(4\pi d^2)$,
where V  is the source  volume in cm$^{3}$,  d is the distance  to the
source $n_e$  and $n_h$ are the  electron and hydrogen  density in the
gas in cm$^{-3}$.  For a gas  temperature of 0.6 keV, it is reasonable
to assume $n_H \sim n_e$. We  then estimate $n_e$ by assuming that the
gas  is   distributed  in   a  sphere  with   radius  $\sim   400$  pc
(corresponding to 2 arcsec, i.e.  the extraction radius for the X--ray
spectrum).} we  obtain $n_e=0.56$  cm$^{-3}$, which is  in agreement
with  the  value  for  $n_e$  we obtain  in  the  electron  scattering
scenario.

We  must point out  that the presence  of the jet  might strongly
affect the properties of the ISM, and the material might well be swept
away along  the line  of propagation  of the jet.   In this  case, the
material  responsible for scattering  may reside  either in  an region
which  surrounds the  jet, or  inside the  jet itself.  In  the latter
scenario, the  electrons of the  jet, which are relativistic  in their
comoving frame  since they emit  synchrotron radiation at least  in the
radio band, may upscatter the nuclear emission via the inverse Compton
process. Electrons  with a Lorentz  factor $\gamma \sim  1000$ (which,
assuming, as a reference value, a magnetic field $B\sim 10^{-4}$ Gauss
emit synchrotron radiation  in the radio band at  $\sim 0.5$ GHz), can
upscatter radio photons of frequency $\nu_s$ to frequencies as high as
$\nu_c \sim \gamma^2 \nu_s$.  For  $\nu_s \sim 1$ GHz and $\gamma \sim
1000$,  we  obtain  $\nu_c  \sim   1  \times  10^{15}$  Hz,  i.e.   UV
emission. Clearly,  more data  at different frequencies  are necessary
for a detailed modeling of synchrotron and inverse Compton radiation.

Although with  the available data  we cannot firmly  distinguish among
the proposed  different scenarios, either of them  are consistent with
the  FR~I--BL Lacs  unification scheme.  When interpreted  in  such a
framework, these observations give us important information on the jet
structure and the physical conditions  in the central tens of parsecs.
In particular,  if the  origin of the  ``A'' component  is synchrotron
radiation from the jet, this  implies that either a velocity structure
in the jet  must be present, or the whole jet  has already slowed down
on the scale of tens  of parsecs.  This provides an important ``link''
to the  evidence for  such stratification in  jets which  have already
been   found   on    larger   \citep{laing99}   and   smaller   scales
\citep{pap3,giovannini99}.

We  argue that  the emission  analogous to  component ``A'',  with its
intrinsic  low surface  brightness, might  be  visible only  in a  few
selected objects.  3C~270  has to be considered as  a special case for
three main reasons: i) its low  distance; ii) the presence of the dark
disk; iii) the particular viewing  angle, at which the ``A'' component
is projected on the dark disk.   We argue that all of these parameters
combine in such way that 3C~270 represents a rather unique opportunity
to observe resolved jet components in the UV regime on the scale of $<
100$pc with current instruments.

The hard power--law  in the X--rays obtained from  the analysis of the
{\it Chandra}  data basically confirms the results  obtained with both
ASCA and  ROSAT.  As already  argued by \citet{fr1_sed},  we interpret
the presence of  a hard X--ray power law component  in the nuclear SED
as due to inverse Compton scattering, since this is indeed expected in
the framework of  the the AGN unification models.   In particular, the
presence  of such  spectral feature  is  a key  characteristic of  low
energy--peaked BL Lac objects  \cite{padovani}, to which 3C~270 should
be unified.

\subsection{Nuclear absorption}

\begin{figure*}
\epsscale{2.0} \plotone{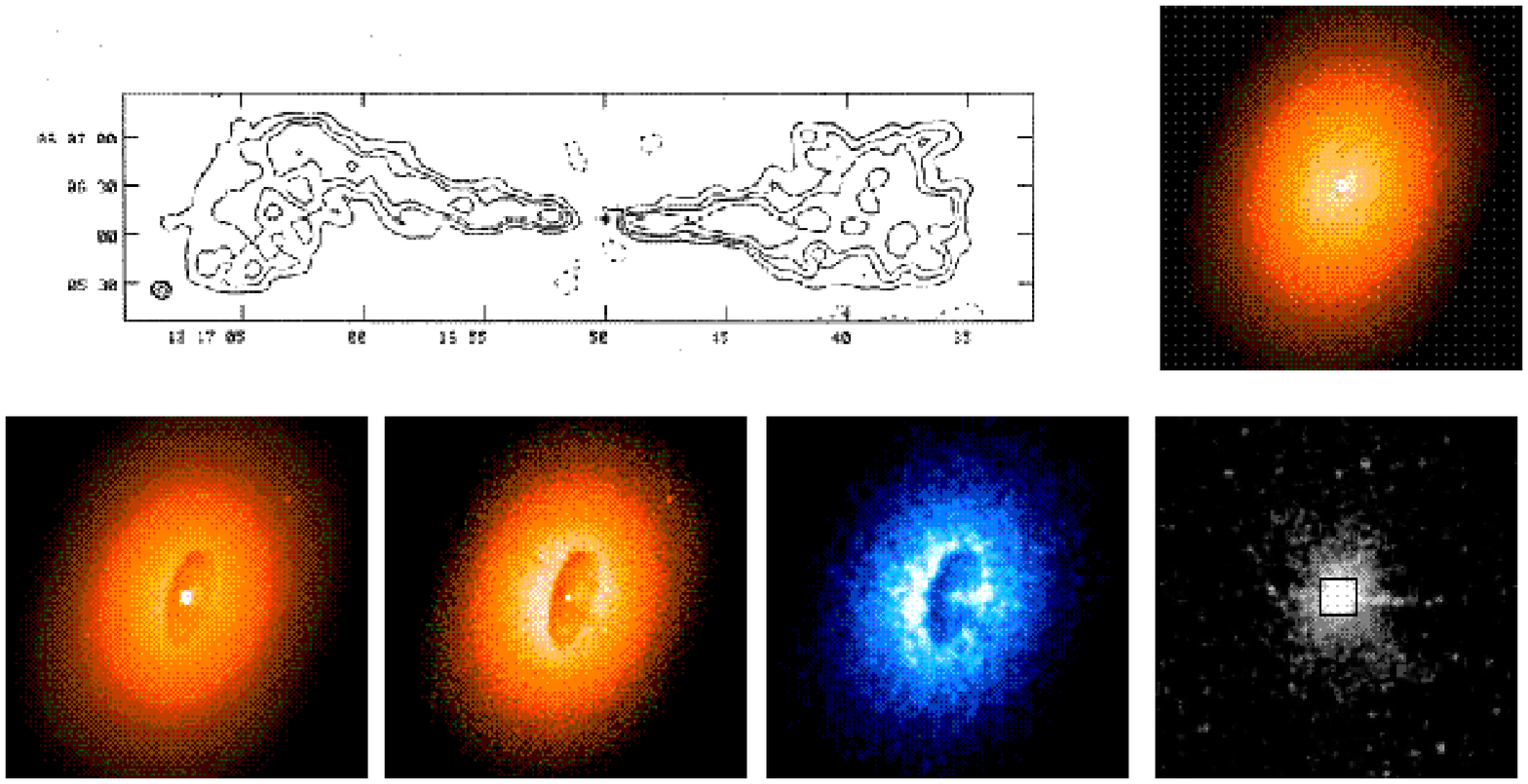}
\caption{A ``picture  gallery'' of 3C~270.  From left  to right (upper
panel) the VLA radio map at 5 GHz as taken from \cite{birkinshaw} (the
asterisk marks the position of the core, which was subtracted from the
map); the infrared image at  1.6 $\mu$m with HST/NICMOS.  Lower panel:
images  at 6700  \AA  ~and 5500  \AA  with HST/WFPC2,  2300 \AA~  with
HST/STIS, and  the X--ray {\it Chandra}  image.  The scale  of the HST
images ($\sim 7^{\prime\prime}  \times 7^{\prime\prime}$), is reported
for reference on the {\it Chandra}  image as a black empty square (see
also other  references in text). North is  up and East is  left in all
images.}
\label{multi}
\end{figure*}

In Fig.  \ref{multi} we show  a ``picture gallery'' of 3C~270, as seen
at different wavelengths.  From left  to right, in the upper panel, we
report the radio map at 5 GHz as taken from \citet{birkinshaw}, images
at 1.6  $\mu$m with HST/NICMOS  \citep{fr1_sed}.  In the  lower panel:
images at  6700 \AA ~and  5500 \AA ~with  HST/WFPC2 \citep{ferrarese},
2300   \AA~  with   HST/STIS  \citep{allen},   and  the   X--ray  {\it
Chandra}/ACIS archival  image.  The HST  images are reproduced  on the
same scale ($\sim 7^{\prime\prime} \times 7^{\prime\prime}$), which is
reported  for reference  on the  {\it Chandra}/ACIS  image as  a black
empty square.   The unresolved nucleus  dominates the emission  of the
central  region of  the  dark disk  in  the 1.6  $\mu$m  and 6700  \AA
~images.   For  shorter   wavelengths  the  nucleus  is  progressively
fainter, and vanishes in the UV.

The absence  of the nucleus, together  with the presence  of a ``jet''
which has  its base  at the location  of the  nucleus, is a  clue that
obscuration is only  confined to the very innermost  region (less than
$\sim  10$pc).   To  investigate   the  properties  of  the  obscuring
structure, we combine the HST data and the X--ray information obtained
with the {\it  Chandra} satellite.  We can evaluate  the absorption of
the dusty  disk by  comparing the counts  in different regions  of the
disk (along  the semi--major  axis) to  the counts in  a small  box of
$5\times 5$  pixels just outside the  disk, $\sim 0.7$  arcsec east of
the nucleus.  We  assume that the galaxy is  optically thin to stellar
light, therefore $1/2$ of the observed counts are produced between the
disk and  the observer, and $1/2$  are produced behind  it.  Along the
line--of--sight to the  disk, only the contribution from  $1/2$ of the
produced  light is  diminished by  the  presence of  dust. Under  this
assumptions  we obtain  for both  the F675W  and the  F547M  images an
optical extinction $A_V=1.2$ magnitudes.   This has to be understood as
a lower  limit to  the disk absorption.   A lower fraction  of stellar
light  produced in  the far  half of  the galaxy  and the  presence of
emission on  the disk (star  forming regions are commonly  observed in
dusty structures of nearby galaxies; e.g.  Allen et al. 2002, Calzetti
et al. 2001) would result in  a higher value of $A_V$. If the fraction
of light from the  back of the galaxy is only $1/3$  of the total, the
value  of $A_V$  increases  to 2.5  (see  Martel et  al.  2000 for  an
extensive discussion on this issue).

A  value of  several magnitudes  for  the nuclear  extinction is  also
compatible with  the analysis of  the spectral energy  distribution of
the  core of  3C~270.  In  \citet{pap_uv} we  have shown  that  in the
framework of  synchrotron emission and assuming  an intrinsic spectral
index $\alpha=1$, the  nucleus vanishes in the UV  image when absorbed
by a column density corresponding to $A_V \sim 3$.  On the other hand,
through the  analysis of  the SED  of 3C~270 in  the framework  of the
unified models,  \citet{fr1_sed} have shown that 3C~270  might be well
unified  to the  low  energy  peaked BL  Lacs  (LBL).  Therefore,  the
intrinsic value  of $\alpha$ for this  source might be  higher than 1,
and this would result in an even lower amount of nuclear absorption.

If we  assume galactic gas--to--dust  ratio to convert $N_H$  to $A_V$
($A_V^X = 5 \times 10^{-22} N_H$),  our results seem to be in contrast
to the  value of $N_H$  as derived from  the analysis of  X--ray data.
Our  value for  the X-ray  absorption  would result  in $A_V=30$  mag.
However, it  is rather  common to find  that the determination  of the
absorbing  column density  from  X--ray information  differs from  the
results  obtained with  other  methods.  Among  Seyfert galaxies,  the
estimate of $A_V$ made both  using absorption features in the infrared
spectrum and evaluating the reddening of optical and infrared emission
lines is typically $0.1-0.5$ times $A_V^{X}$ \citep{maiolino,granato}.
A similar  result has also  been recently found by  \citet{marconi} in
the  case of  the closest  FR~I radio  galaxy, Centaurus  A,  in which
absorption towards the nucleus can  well be provided by the kpc--scale
dust lane.  Interestingly,   \citet{mamaol} have proposed that the
excess  of X-ray  absorption compared  to the  lower extinction  in the
optical and  infrared can  be accounted for  by the presence  of large
grain dust  which have small extinction  efficiency.  Another possible
interpretation is  that in 3C~270 the absorbing  structure is ``free''
from dust.

Therefore, the  observations of 3C~270  fit the scenario in  which the
nuclei  of FR~I radio  galaxies are  not obscured  by the  presence of
optically and  geometrically thick tori.  The  observed (rather small)
amount of  absorption in 3C~270 can  be well provided  by the extended
dusty disk seen almost edge--on.

\section{Summary and conclusions}
\label{conclusions}

We  have  analyzed  and  discussed  HST  and  {\it  Chandra}  data  of
3C~270.  The HST/STIS UV  image shows  a jet--like  feature (component
``A'')  originating from  the location  of  the nucleus,  at the  same
position angle  as the radio and  X--ray jet.  The  nucleus is clearly
visible in  the IR and optical images,  and it is not  detected in the
UV.  On  the other hand,  the emission of  the jet-like feature  is so
low, that  it cannot  be detected  in the optical  images, due  to the
higher amount of stellar emission in that band compared to the UV. The
color  image  UV--V  shows that  the  jet  is  indeed bluer  than  the
underlying stellar emission, while the radio--to--UV spectral index is
consistent with other synchrotron jets.

We propose two  basic scenarios for the origin  of component ``A'': i)
non-thermal synchrotron emission from a slowly moving (possibly mildly
relativistically)  component of  the relativistic  jet;  ii) scattered
radiation. In  case i),  this would argue  for the presence  of mildly
relativistic  motions in  the  jet on  the  scale of  tens  of pc,  in
addition to the already achieved  evidence for such motions on smaller
(VLBI)  and larger (VLA  and X--rays  jets) scales.   In case  ii) the
ultimate origin for the emission may well reside in radiation from the
base of the jet, which is scattered into our line of sight, confirming
that 3C~270  is indeed a misoriented  BL Lac.  To  test such pictures,
co--spatial radio  data are needed  in order to identify  the possible
radio counterpart  of the UV  jet--like feature and thus  constrain the
broad--band  spectral  index  and  the  emission  mechanism.   Another
possible test  for discriminating between the  different scenarios can
be provided by optical (and radio) polarimetric observations.  In case
of  scattered  radiation  we  expect  the radiation  to  be  polarized
perpendicularly  to  the  jet   direction.   In  case  of  synchrotron
emission, the  polarization properties  must be strictly  connected to
what is observed in the radio jet.

The X--ray  spectrum of the nucleus  shows a flat  ($\alpha \sim 0.3$)
power--law  component,  absorbed  by  a  column  density  $N_H=6\times
10^{22}$ cm$^{-2}$.   The apparent discrepancy between  this value and
the  low amount  of absorption  estimated in  the optical,  is however
often  found in  other AGNs,  such  as Seyfert  galaxies. The  nuclear
structure of 3C~270 appears to be in agreement with that of other FR~I
radiogalaxies: a standard  optically and geometrically thick obscuring
torus is  not required  to account for  the observed  properties.  The
origin  for  the  discrepancy  between  the amount  of  absorption  as
estimated through  different methods is still to  be fully understood.
The power--law component in the X--rays is a further indication of the
presence of a  BL Lac in its nucleus.  We confirm  that the flat slope
observed for  such a  component can be  interpreted as due  to inverse
Compton   emission,   analogously  to   what   is   observed  in   low
energy--peaked BL Lacs.

\begin{acknowledgements}
The authors wish  to thank E. Trussoni, M. Murgia  and G. Risaliti for
useful comments.
\end{acknowledgements}

\clearpage


\clearpage

\clearpage

\end{document}